\documentclass[twocolumn]{aastex62}
\usepackage{hyperref}
\usepackage{bm}
\usepackage{multirow}

\newcommand{\Athena}{\textsc{Athena\scriptsize ++ }}

\def\avg#1{\langle#1\rangle}
\def\MSUN{\rm M_{\odot}}
\received{December 2019}
\revised{April 2020}
\accepted{April 2020}
\submitjournal{ApJ}
\shorttitle{Clumpy outflows}
\shortauthors{Dannen et al.}
\begin{document}
\title{Clumpy AGN outflows due to thermal instability}
\correspondingauthor{Randall Dannen}
\email{randall.dannen@unlv.edu}
\author[0000-0002-5160-8716]{Randall C. Dannen}
\affiliation{Department of Physics \& Astronomy \\
University of Nevada, Las Vegas \\
4505 S. Maryland Pkwy \\
Las Vegas, NV, 89154-4002, USA}
\author[0000-0002-6336-5125]{Daniel Proga}
\affiliation{Department of Physics \& Astronomy \\
University of Nevada, Las Vegas \\
4505 S. Maryland Pkwy \\
Las Vegas, NV, 89154-4002, USA}
\author[0000-0002-5205-9472]{Tim Waters}
\affiliation{Department of Physics \& Astronomy \\
University of Nevada, Las Vegas \\
4505 S. Maryland Pkwy \\
Las Vegas, NV, 89154-4002, USA}
\author{Sergei Dyda}
\affiliation{Institute of Astronomy \\ Madingley Road, Cambridge CB3 0HA, UK}
\begin{abstract}
One of the main mechanisms that could drive mass outflows on parsec scales 
in AGN is thermal driving. The same X-rays that ionize and heat the plasma 
are also expected to make it thermally unstable. Indeed, it has been proposed 
that the observed clumpiness in AGN winds is caused by thermal instability (TI). 
While many studies employing time-dependent numerical simulations of AGN 
outflows have included the necessary physics for TI, none have so far managed 
to produce clumpiness. Here we present the first such clumpy wind simulations 
in 1-D and 2-D, obtained by simulating parsec scale outflows irradiated 
by an AGN. By combining an analysis of our extensive parameter survey
with physical arguments, we show that the lack of clumps in previous numerical 
models can be attributed to the following three effects:
(i) insufficient radiative heating or other physical processes that prevent 
the outflowing gas from entering the TI zone; 
(ii) the stabilizing effect of stretching 
(due to rapid radial acceleration)
in cases where the gas enters the TI zone; 
and (iii) a flow speed effect: in circumstances where stretching is inefficient, 
the flow can still be so fast that it passes through the TI zone too quickly 
for perturbations to grow. Besides these considerations, we also find that 
a necessary condition to trigger TI in an outflow is for the pressure ionization 
parameter to decrease along a streamline once gas enters a TI zone.
\end{abstract}
\section{Introduction} \label{sec:intro}
Thermal instability \citep[TI,][]{Field65} was long ago recognized as a viable mechanism 
for producing multiple phases in AGN winds \citep[e.g.][]{DN79,KV84,SVS85}.  
Such winds are observed, for example, in some Seyfert galaxies, where the UV and X-ray 
absorbers have similar velocities, strongly suggesting that 
very different ions are nearly cospatial and therefore that different temperature 
regions coexist \citep[e.g.,][and references therein]{SH97,Cetal99, Getal03,
Letal13, Eetal16,Fetal17, Metal17}.  

While TI is well understood in a local approximation \citep[e.g.,][]{Balbus95,WP19},
it has proven challenging to quantitatively model clump formation in a dynamic flow.
Only in recent years has it become clear how the in situ production of multiphase gas 
can be triggered using \textit{global} time-dependent hydrodynamical simulations -- 
and only in the context of accretion flows \citep[][MP13 hereafter]{Barai12,GRO13,TOM13,MP13} 
or stratified atmospheres \citep[e.g.,][]{MSQ12,SMQ12}.  
In an outflow regime, previous work has focused on highlighting 
the  importance of considering the effects of clumpiness, but only on a qualitative basis 
\citep[e.g.,][]{Nayakshin14, Elvis17}.  

Here, we present the first global simulations of  
an outflow that is multiphase due to TI.  Specifically, we show that there exists 
a range of the so-called hydrodynamic escape parameter (HEP),
small yet relevant, for which the outflow can develop regions with significant 
over- and under-densities and that the over-densities 
can survive their acceleration over a relatively large distance.   
We identify the physical effects that contribute to making the HEP range
small and thus explain why we and others have not yet seen a clumpy outflow 
in any previously published results from the simulations of thermally driven outflows, 
neither in 1- nor 2-D simulations 
\citep[e.g.,][]{Wetal96, PK02, Letal10, HP15, Detal17, WP18}. 
We detail our numerical methods in \S{\ref{sec:num-meth}}.  
We present the results from our calculations in \S{\ref{sec:results}} 
and a discussion in \S{\ref{sec:discussion}}.
\par

\section{Numerical Methods} \label{sec:num-meth}
We employ the magnetohydrodynamics code \Athena \cite{Stone20}
to solve the equations of non-adiabatic gas dynamics
\begin{equation} \label{eq:hydro-matter}
\frac{\partial \rho}{\partial t} + \nabla\cdot (\rho {\bf v}) =0
\end{equation} 
\quad
\begin{equation} \label{eq:hydro-mom}
\frac{\partial \rho {\bf v}}{\partial t} 
+ \nabla\cdot (\rho {\bf v}{\bf v} + {\bf P}) = 
-\rho \nabla \Phi + {\bf F}_{\text{rad}}
\end{equation} 
\quad
\begin{equation} \label{eq:hydro-en}
\frac{\partial E}{\partial t} + \nabla \cdot [(E+P){\bf v}] =
-\rho{\bf v} \cdot \nabla \Phi - \rho \mathcal{L} + 
 {\bf v}\cdot {\bf F}_{\text{rad}}
\end{equation} 
where $\rho$ is the fluid density, ${\bf v}$ is the fluid velocity, ${\bf P} = p\, {\bf I}$ with $p$ the gas pressure and ${\bf I}$ the unit tensor, 
$\Phi=-GM_{\rm BH}/r$ is the gravitational potential due to a black hole 
with mass $M_{\rm BH}$, $E = \rho \mathcal{E} + 1/2 \rho |{\bf v}|^2$ 
is the total energy with $\mathcal{E} = (\gamma -1)^{-1}p/\rho$ 
the gas internal energy, ${\bf F}^{\text{rad}}={F}^{\text{rad}}\, \hat{\bm r}$ is the radiation force, and $\mathcal{L}$ is the net cooling rate.  All of our calculations assume $\gamma = 5/3$.  
Although we have computed models with the radiation force
due to both electron scattering and line-driving,
in this Letter we present only models with the force due
to electron scattering, i.e., 
${F}_{\text{rad}} = {GM_{\rm BH}\rho\,\Gamma}/r^{2}$, where
$\Gamma=L/L_{\rm Edd}$ is the Eddington fraction.
Irradiation is thus assumed to be due to a point source, 
as is appropriate when considering parsec scales in relation to 
the X-ray coronae and the UV emitting regions of AGN disks. 

\begin{figure}[htb!]
 \centering
 \includegraphics[width=0.95\columnwidth, height=0.45\textheight]{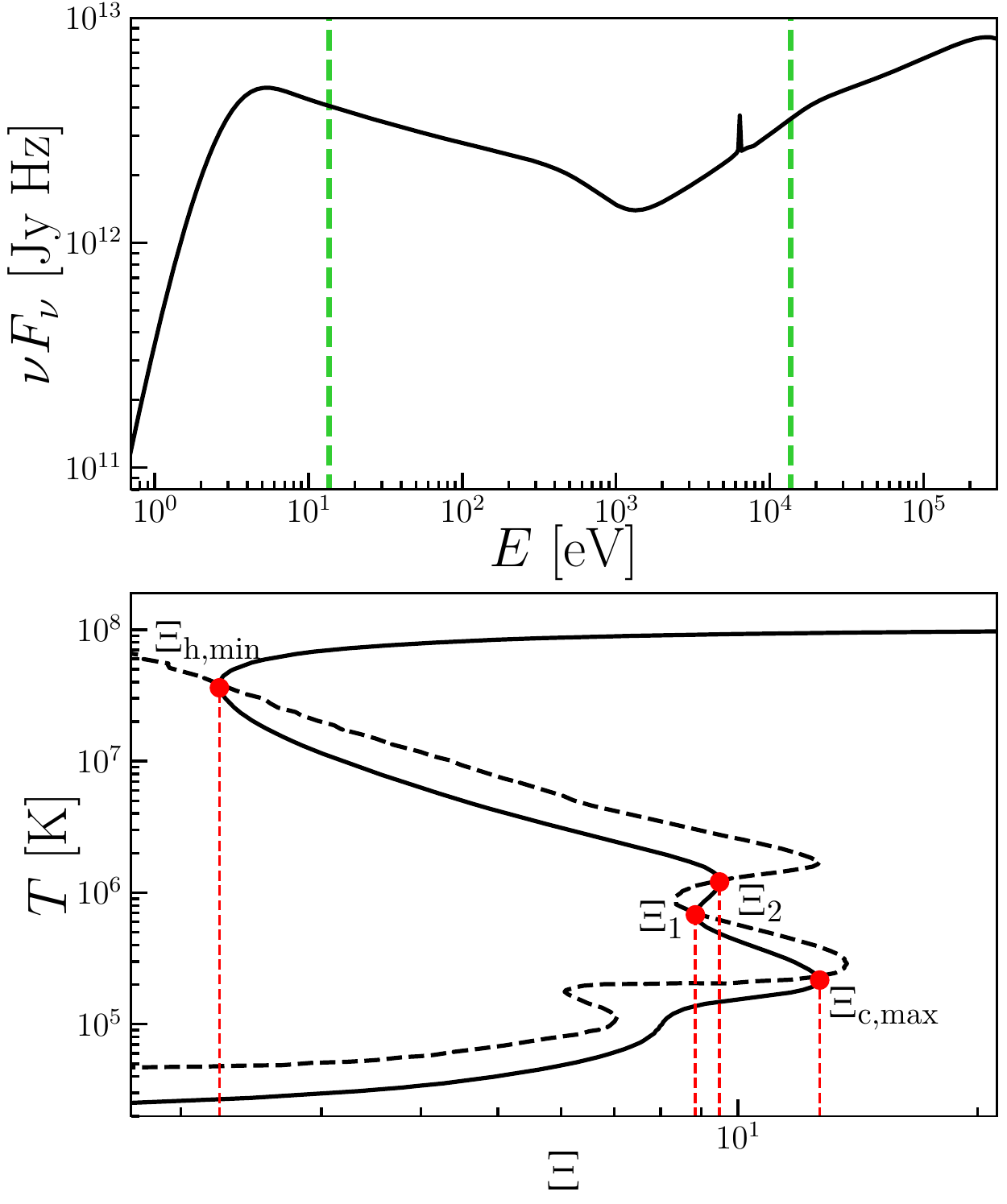}
 \caption{
{\it Top panel:} 
SED intrinsic to NGC 5548, as determined by \cite{Metal15}. 
The energy range used to define $\xi$ is marked by the two vertical lines. 
This SED is relatively flat and has $L_X/L \approx 0.36$ and mean photon energy $\avg{h \nu} =k_{\rm B} T_{\rm X}=34.8$~keV (corresponding to Compton temperature $T_{\rm C}=T_{\rm X}/4=1.01 \times 10^8$~K). 
{\it Bottom panel:} Associated S-curve and Balbus contour (the solid and dashed lines, respectively).
Red dots mark the points $\Xi_{\rm c,max}$, $\Xi_{1}$, $\Xi_{2}$, and $\Xi_{\rm h,min}$.  Note that  $\Xi_{\rm c,max}$ ($\Xi_{\rm h,min}$) denotes the last (first) stable point on the ``cold phase branch'' (``Compton branch'') of the S-curve.
The SED-dependent conversion to the other common ionization parameter $U = (\Phi_{\rm H} / c)/n$ (with $\Phi_{\rm H}$ the number density of \ion{H}{1} ionizing photons) is $U\approx \xi / 42$, so 
$[\log(\Xi_{\rm c,max}), \log(\xi_{\rm c,max}), \log(U_{\rm c,max})] = [ 1.10, 2.15, 0.53 ]$. 
}
 \label{fig:sed-s-curve}
\end{figure}

\begin{table*}
\centering
\begin{tabular}{ c | c c c c c c c | l l }
\multirow{2}{*}{Model} & \multirow{2}{*}{HEP} &       $r_{0}$  &               $\rho_{0}$ &  $t_{\rm sc,0}$ & \multirow{2}{*}{$t_{\rm cool} / t_{\rm sc}$} &              $\avg{v}$ &        $\avg{\dot{M}}$ &  \multicolumn{2}{c}{Comment}        \\ 
                       &                      & [$10^{18}$ cm] & [$10^{-18}$ g cm$^{-3}$] & [$10^{12}$ s] &                                              & [$10^{7}$ cm s$^{-1}$] & [$10^{24}$ g s$^{-1}$] & 1x runs      & (8x runs)  \\ \hline
                     A & 13.3                 &           1.01 &                     2.68 &           3.4 &                                  0.47 (0.48) &              6.4 (6.0) &              1.2 (1.2) & steady       & (steady) \\
                     B & 11.9                 &           1.13 &                     2.15 &           3.9 &                                  0.54 (0.49) &              3.4 (3.5) &              3.2 (3.3) & unsteady     & (unsteady) \\
                     C & 9.1                  &           1.48 &                     1.26 &           5.0 &                                  0.41 (0.42) &              2.5 (2.3) &              5.9 (6.0) & unsteady     & (unsteady) \\
                     D & 8.1                  &           1.67 &                     0.98 &           5.7 &                                  0.41 (0.41) &              2.4 (2.2) &              6.6 (6.7) & quasi-steady & (unsteady) \\
                     \end{tabular}
\caption{\small {Summary of key parameters and results.  HEP is the hydrodynamic escape parameter.  The inner radius is derived from the choice of HEP as $r_0 = \left(1 - \Gamma \right) GM_{\rm BH}\text{HEP}^{-1} c_{s,0}^{-2}$ (see Eq.~\ref{eq:hep}), while the density at the base follows from the definition of $\xi$ as $\rho_0 = \mu m_{p} L_{\rm X}\xi_{0}^{-1} r_{0}^{-2}$.  The sound crossing time is defined as $t_{\rm sc,0} = (r_{\rm out} - r_{0}) / c_{\rm s,0}$.  
The ratio $t_{\rm cool} / t_{\rm sc}$ is given at the location where $\Xi = \Xi_{\rm c,max}$.
The average mass flux and velocity through $r_{\rm{out}}$ are shown as $\avg{\dot{M}}$ and $\avg{v}$.
Comment columns denote the state of the flow at late times.  Values/comments in parenthesis denote results for the 8x-resolution runs.
}}
\label{table:summary}
\end{table*}

For the unobscured AGN SED of NGC 5548 from \cite{Metal15}, shown in the top panel of 
Fig.~\ref{fig:sed-s-curve}, 
D17 have tabulated the net cooling rate $\mathcal{L} = \mathcal{L}(T,\xi)$, 
which is function of the gas temperature, $T$, and the ionization parameter, 
defined as
$\xi = L_{\rm X} /n r^2$
where $L_{\rm X}$ is the luminosity integrated between 1-1000~Ry and 
$n = (\mu m_p)^{-1}\rho$ is the gas number density (with $m_p$ the proton mass; we set $\mu = 0.6$ in this work).
The $\xi$ dependence on distance plays an important role in determining 
the thermal stability of the flow, as discussed in detail in \S\ref{sec:discussion}.
Also, the ratio between the hard and soft X-ray energy bands
is an important characteristic affecting properties of TI \citep[e.g.,][]{KMc82,K99, Metal15, Detal17}. 

\par
The solid line in the bottom panel of Fig.~\ref{fig:sed-s-curve} is
the S-curve corresponding to this SED,
i.e. the contour where $\mathcal{L}(T,\Xi)=0$, where $\Xi \equiv  p_{\rm rad}/p=\xi/(4\pi ck_{\rm B}T)$ is the pressure ionization parameter, with $p_{\rm rad}$ the radiation pressure (equal to $F_{\rm X}/c$ in our models). 
For the adopted SED, gas is unstable by the isobaric criterion for local TI 
in the two zones where the slope of the S-curve is negative (these are 
the locations where Field's criterion for TI, 
$[\partial \mathcal L/\partial T]_{p} <0$, is satisfied).
Gas is actually thermally unstable everywhere left of the dashed line (hereafter referred to as the Balbus contour), 
as this region of parameter space satisfies Balbus' generalized criterion for TI, 
$[\partial (\mathcal L/T)/\partial T]_{p} <0$ \citep{Balbus86}. 
Points corresponding to the maximum value of $\Xi$  
on the cold stable branch of the S-curve
(i.e, for $\log~\Xi_{\rm c, max} = 1.10$)
are dynamically the most significant, as they mark the entry into a cloud formation 
zone and dramatic changes in the flow profiles can occur there.
\par


We present the results of 1-D models run at both medium and high resolution, 
all using a uniform grid spacing [$\Delta r \approx (r_{\rm out} - r_{\rm 0})/N_r$].
The standard resolution runs (Models A-1x, B-1x, C-1x, \& D-1x) have $N_r = 552$,
chosen such that gradient scale heights $\lambda_q \equiv q/|dq/dr|$ (where $q$ 
is any hydrodynamic variable) are adequately resolved with 
$\lambda_q/\Delta r \approx 3$ for $r \lesssim 1.1\,r_0$ and 
$\lambda_q/\Delta r > 10$ at all other points in the wind except in 
the vicinity of the location where $\Xi = \Xi_{c,\rm{max}}$, where this 
ratio dips to about 1.  The high resolution runs (Models A-8x, 
B-8x, C-8x, \& D-8x) have $N_r = 8\times552$, nearly an order of 
magnitude higher resolution everywhere.
The other relevant length scale to resolve is that of the clumps, 
which have a size on the order of the local cooling length scale 
$\lambda_{\rm{cool}} \equiv c_s\, t_{\rm{cool}}$, where
$t_{\rm cool}=\mathcal{E} (n \,\mathcal{C} / m_p)^{-1}$ is the cooling time
while $\mathcal{C}$ is the cooling rate in units of 
${\rm erg\, cm^{3}\, s^{-1}}$.  For $\gamma = 5/3$,
\begin{equation} \label{eq:l_cool}
\lambda_{\rm{cool}} \approx 3 \times 10^{16}\,T_5^{3/2}\, n_3^{-1}\, \mathcal{C}_{23}^{-1}\: \rm{cm},
\end{equation}
where $T_5 = T/10^5\,\rm{K}$, $n_3 = n/10^3\,\rm{cm^{-3}}$, $\mathcal{C}_{23}= \mathcal{C}/10^{-23}\,\rm{erg\,cm^3\,s^{-1}}$, and $T_5 = n_3 = C_{23} = 1$ are characteristic values of the clumps.  With $\Delta r \approx 2\times 10^{15}\,\rm{cm}$ for our `-1x' runs, this length scale is adequately resolved. 

We apply outflowing boundary conditions at the inner 
and outer radii with the density fixed at the innermost active grid point 
to the value of $\rho_0$ in Table~\ref{table:summary}.  The initial conditions are 
a simple expanding atmosphere with $\rho = \rho_0\, (r/r_0)^{-2}$ and a $\beta$-law 
velocity profile, $v = v_{\rm esc} \sqrt{1 - r_{0}/r}$, 
where the ``0'' subscript is used to denote values at the inner radius 
of the computational domain, $r_0$.  
The pressure profile is set according to the temperature, 
chosen so that the gas lies along the S-curve, having a value 
$\Xi_0$ at $r_0$ (see \S\ref{sec:results}).  
\par
The 2-D model uses a $256\times256$ grid in $r$ and $\theta$, with logarithmic 
spacing in $r$ such that $dr_{i+1} = 1.01~dr_{i}$ and uniform spacing in $\theta$ 
from 0 to $\pi$.  At the inner boundary we assumed a density profile 
$\rho = \rho_0[1 + 0.001\sin(2\theta)]$ to break spherical symmetry.
We apply reflecting boundary conditions at 0 and $\pi$.
\begin{figure*}[htb!]
  \centering
  \includegraphics[width=0.96\textwidth]{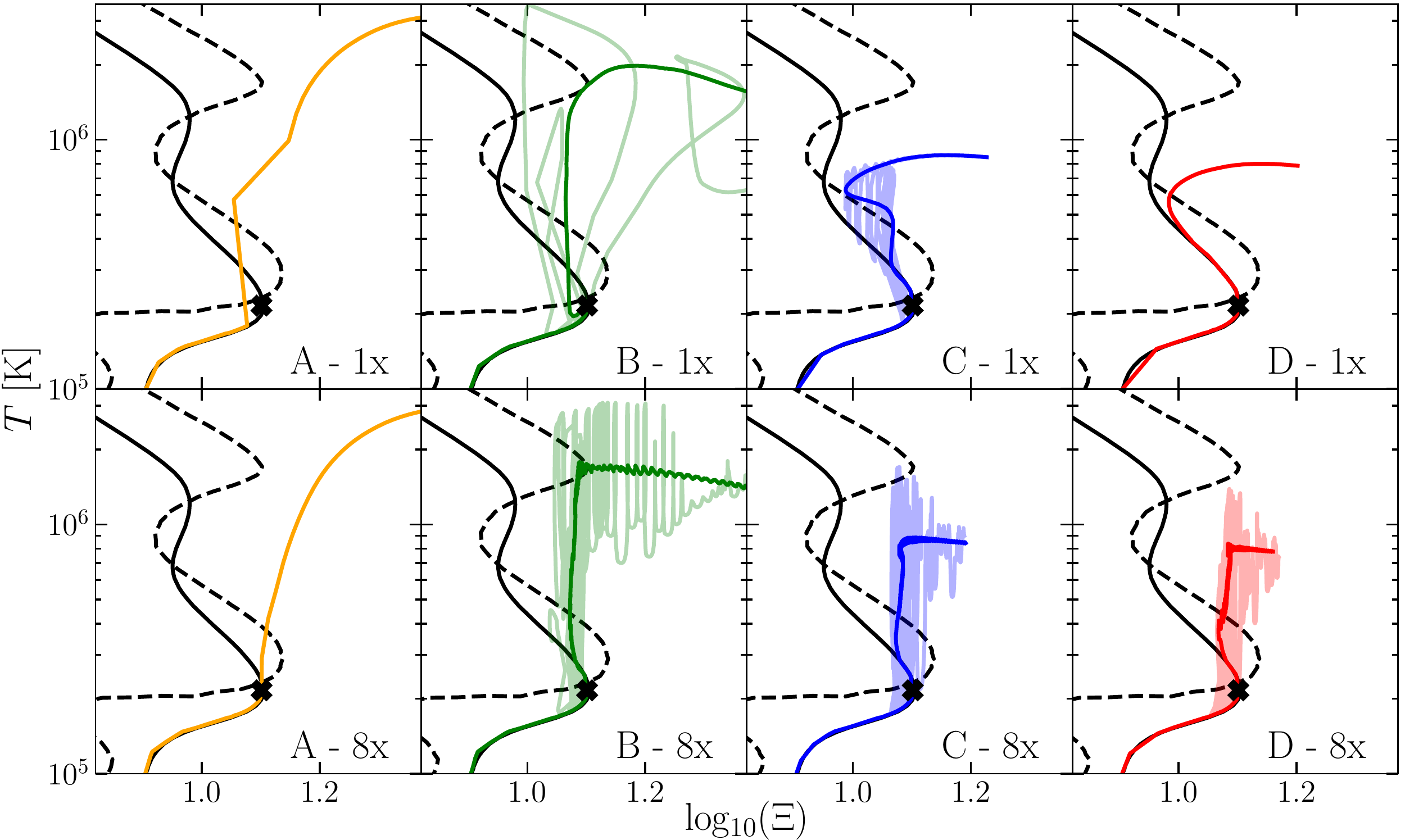}
  \caption{Phase diagram comparison of our four 1-D models showing $T$ as function of $\Xi$ in relation to the thermal equilibrium curve (black solid line) and the Balbus contour (black dashed line).  The colored solid lines display time-averaged solutions and the less opaque lines a single snapshot at the end of the simulation  (at $t > 3\,t_{\rm sc,0}$).  The top row presents our standard resolution (`-1x') runs
   and the bottom row our high (`-8x') resolution runs.  See 
   \S\ref{sec:results} and \S\ref{sec:discussion} for details about the dynamics of each model.}
  \label{fig:1d-summary-phase}
\end{figure*}
\section{Results}\label{sec:results}
For a given $M_{\rm BH}$, just three parameters govern our solutions:
$\Gamma$, $\Xi_0$, and the HEP, which sets the strength of thermal driving. 
We define HEP as the ratio of effective gravitational potential 
and thermal energy at $r_0$,
\begin{equation}\label{eq:hep}
    \text{HEP} = \frac{GM_{\rm BH}(1 - \Gamma)}{r_0 c_{s,0}^2}.
\end{equation}
We only present models with $\Xi_0 = 3$, $\Gamma = 0.3$, and 
$M_{\rm BH}=10^{6}~\MSUN$, as HEP is the main governing parameter.  
We note, however, that the choice of $\Xi_{0}$ is not unimportant, e.g.,
selecting one too large can cause the flow to miss relevant regions.  
The dependence of stable wind solutions on $\Gamma$ for various SEDs was explored in 1-D by D17.
In Table~\ref{table:summary}, we list model parameters and summarize gross outflow properties. 

After examining over one hundred 1-D simulations,
with the standard resolution,
that span this parameter space,
we arrived at four qualitatively different 1-D 
wind solutions that capture the general behavior seen across all runs. 
These four cases illustrate how the stability of the outflow depends 
on the HEP (and the numerical resolution). We chose the values of HEP for our
models~A and D 
such that they closely bracket the parameter space leading to transonic, 
clumpy winds (here represented by models B, C, and D-8x). \par
For each model, Fig.~\ref{fig:1d-summary-phase} shows the models tracks on the phase diagram while Fig.~\ref{fig:1d-summary-prims} shows radial profiles of $\rho$, $v$, $T$, and $\Xi$.  The phase diagrams reveal that all solutions pass through the lower TI zone but only models~B and C 
with the standard resolution
actually trigger TI to become unsteady.  
Additionally, two general trends are evident: 
(i) the range of $\Xi$ decreases with ${\rm HEP}$ and 
(ii) for large $\Xi$, the wind is not in thermal equilibrium; 
the wind temperature is nowhere near that of the stable Compton 
branch on the S-curve (the maximum temperature shown 
in these plots is more than an order of magnitude lower than $T_C$). 
The first trend is due to the fact that for thermal winds, 
the velocity is a decreasing function of ${\rm HEP}$, and 
by mass conservation for radial flows, $\xi \propto v$ in a steady state 
(i.e. $v \propto  1/n\, r^2$, as is $\xi$).
Trend (ii) is due to adiabatic cooling (see D17), but note that 
the highest temperatures reached in all four cases occupy thermally 
stable regions of the ($T$--$\Xi$)-plane (namely, regions to the right 
of the Balbus contour).  
We now examine the dynamics of these solutions in detail to understand 
why Models~B, C, \& D-x8 are clumpy, while A \& D-x1 are not.  

\subsection{Unsteady, clumpy wind solutions}
\label{sec:clumpyruns}
\begin{figure*}[htb!]
  \centering
  \includegraphics[width=\textwidth]{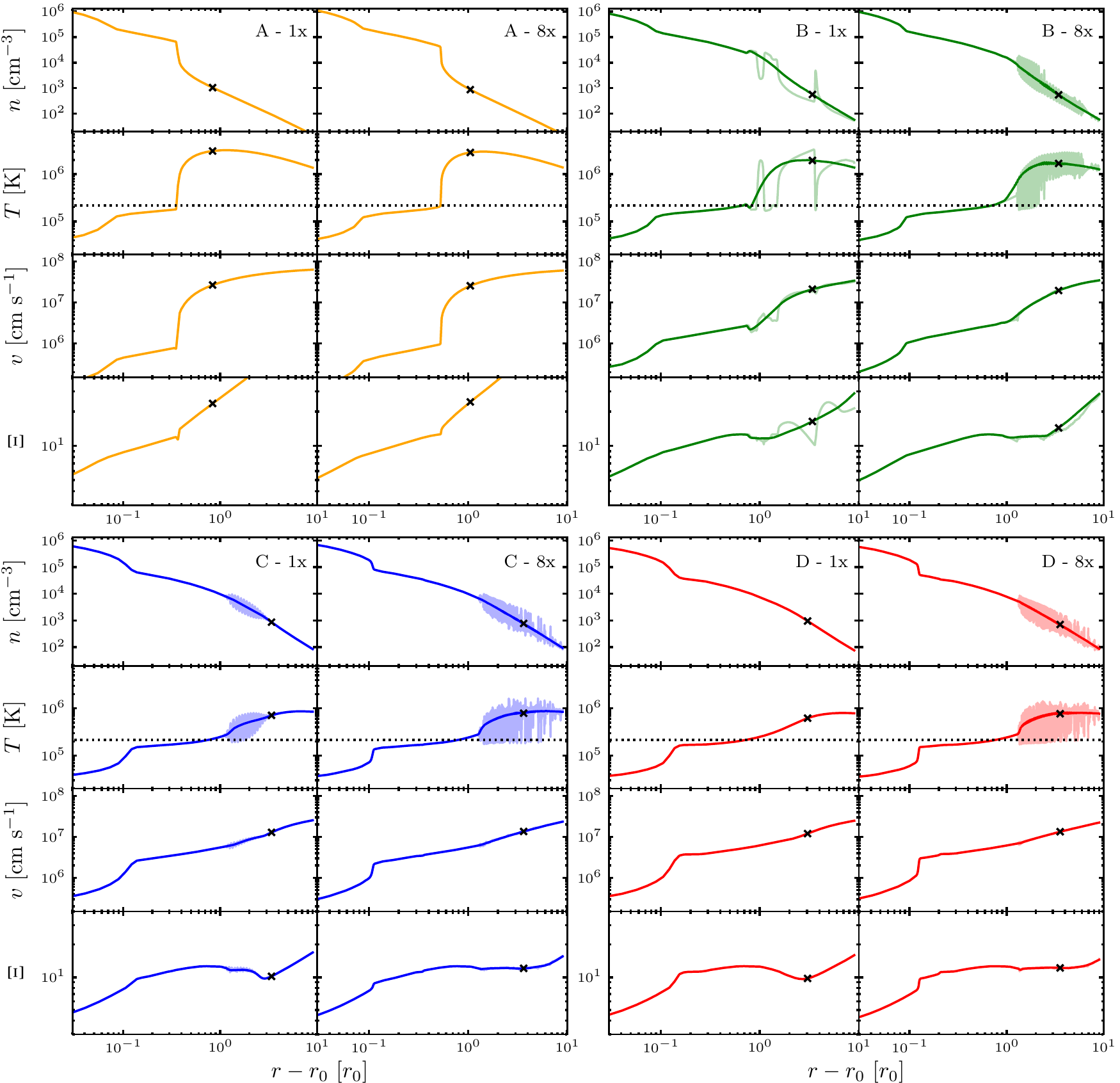}
  \caption{Spatial flow profiles of our four 1-D models run at both medium (`-1x' runs, 
  the first and third columns of panels) and high (`-8x' runs, the second and fourth column
  of panels) resolution. As in Fig.~2, fully opaque curves show time-averaged profiles, while less opaque ones are snapshots from the end of the simulation.   
  The $\mathrm{X}$'s mark sonic points of the time-averaged solutions.  
  In the temperature panels, the dotted black line shows where 
  $T=T(\Xi_{\rm c,max})$.  Animations of these runs are downloadable at
  \url{https://doi.org/10.5281/zenodo.3739603} or watchable at
  \url{http://www.physics.unlv.edu/astro/clumpywindsims.html}.
   }
  \label{fig:1d-summary-prims}
\end{figure*}

A basic requirement for TI to operate once gas enters the TI zone 
is for it to stay there long enough for initially small perturbations 
to grow.  In terms of timescales, the flow must satisfy $t_{\rm{cool}} < t_{\rm{sc}}$ on global scales, which it does (see Table~\ref{table:summary}).  However, Fig.~\ref{fig:1d-summary-phase} shows that for the flow to remain in the TI-zone, it must also be the case that $\Xi$ \textit{not} increase downstream, the normal tendency in disk winds, at least in the absence of magnetic field pressure \citep{HP15}. 
This normal $\Xi$-scaling is especially obvious in the 1-D radial winds 
studied by D17 that had constant radiation flux. In such winds,  
$\Xi \propto 1/p$, and $p$ tends to decrease radially outward 
in outflows, the necessary condition for the pressure gradient 
to overcome gravity.
The increase in $\Xi$ will therefore cause the gas to quickly leave 
the TI zone (i.e. to cross to the right of the dashed lines in 
Fig.~\ref{fig:1d-summary-phase} once reaching $\Xi_{\rm c, max}$).  
What we see in models~B, C, and D, however, 
is that $\Xi$ can decrease outwards through the TI-zone (see the bottom panels in Fig.~\ref{fig:1d-summary-prims}).  Viewed in terms of the phase diagrams, gas not only enters and stays in the TI zone, but even crosses through a large portion of this zone. 
That a necessary condition for a clumpy wind solution is for the gas 
pressure profile to be such that it leads to a decrease of $\Xi$ within 
a TI zone is the critical insight of this work. 
The criterion that $\Xi$ decrease outward is hard to satisfy 
(see \S\ref{sec:discussion}), explaining why TI is absent in 
the thermally driven wind models studied in the past.

In model~B, the inner gas is just slightly less gravitationally bound 
than in model~A (HEP of 11.9 compared to 13.3), yet this is enough 
for the flow to be twice as fast at small radii and to follow the
S-curve all the way to the TI zone.
As the flow enters this zone, we see formation 
of an initially thin layer where the temperature increases rapidly 
(see the locations marked by dotted lines in Fig.~\ref{fig:1d-summary-prims}). 
This heating is radiative and is related to the gas being thermally unstable.
We note that the heating is localized, namely, the gas immediately  
interior and exterior of this hot layer is relatively cool and dense, 
as it is gas located on or just above the upper branch of the cold phase.
The separation of this cooler downstream flow that has entered the TI zone 
from the cold phase upstream flow by the hot layer is the origin of 
a dense clump that, in the animations of these runs (see Fig.~\ref{fig:1d-summary-prims} for a link), appears as an ejection of a layer of cold gas.\par

We reiterate that this is not clump formation from a condensation as in 
classical TI, as the role of TI here is primarily to form a hot layer.  
This layer is initially cool, lying above a stable cold region 
and below an unstable and condensing cool region, but there is very little 
`room' in the ($T$--$\Xi$)-plane for gas to condense, 
while there is a lot of room for it to heat.  
With time, the hot layer expands and 
pushes the colder, dense material outward. The ram pressure increases somewhat
the density of the cold gas but it is not the cause of the over-densities.
It would be more appropriate to view this process
as the separation of layers of the atmosphere 
rather than the
condensation of cool clumps.
\par


We note that TI operates here under nearly isobaric conditions
(the cold and hot phases in Fig.~\ref{fig:1d-summary-phase} are 
connected by nearly vertical lines at a given radius).  
In particular, the size of the perturbation 
that grows, as well as the resulting clumps,
are both smaller than the pressure scale height, and therefore they have 
nearly constant pressure.  Despite having $t_{\rm{cool}} < t_{\rm{sc}}$ 
globally (i.e. on the scale of $\lambda_p$) as mentioned above, the 
requirement for isobaricity ($t_{\rm{sc}} < t_{\rm{cool}}$) is satisfied locally.

Finally, while it is not obvious from Figs.~\ref{fig:1d-summary-phase} 
and~\ref{fig:1d-summary-prims}, we found that generally as the flow accelerates,
$\xi$ increases, even for the cold phase gas. 
This leads to heating and expansion of the colder clumps,
eventually causing them to enter the hot phase.   
In model~C-1x, this expansion process operates over a relatively small radial range and is well displayed in the figures as a decrease in the density and temperature variations with distance.
\par

\subsection{Smooth wind solutions passing through a TI zone}
Models with HEP higher than that for model~B evolve toward 
a steady, transonic and stable solution of a Compton heated wind, 
while all other models 
are \emph{unstable} (and hence unsteady) transonic solutions 
of a thermally driven wind heated mainly by photo-absorption
(the temperature at the sonic point and even of the hottest gas 
is less than $10^{6}$~K). 
Model~A resembles the cases that we saw in the past (e.g., D17). 
At small radii, the gas is subsonic with a density profile close to that 
of a hydrostatic equilibrium solution 
for a temperature profile tracing the cold branch of the S-curve. 
At $r\approx1.4\, r_{\rm 0}$, where $\log~\Xi\simeq1.1$, the flow undergoes
runaway heating (see the temperature panels in Fig.~\ref{fig:1d-summary-prims}
for Model B). The heat input is relatively large and it results 
in a rapid flow acceleration, the wind becoming supersonic 
at $r\approx 2\,r_{\rm 0}$.
Similarly to the cases described in D17, the increase in $v$ is 
so significant that the adiabatic cooling becomes faster than 
the radiative heating and the flow does not evolve along 
the S-curve at large radii. %
\par
Two dynamical effects suppress TI in solutions with relatively high HEP: 
1) stretching (i.e. radial flow acceleration that causes rapid 
expansion of fluid elements) operating on the time scale 
$\tau_{\rm s}=1/|\partial v/ \partial r|$ and 
2) high velocity.  The latter causes a gas parcel to `fly through' 
the TI zone on the dynamical time scale 
$\tau_{\rm d}=|r/v|$, which is shorter than the maximum growth rate of TI, 
$\tau_{\rm TI}$.
These two effects are the main reason for an accelerating flow 
to be thermally stable despite satisfying Balbus' generalized 
instability criterion (see discussions of this point in MP13
and \cite{HP15} in the context of accretion flows and
of thermal disk winds, respectively).

\begin{figure*}[htb!]
  \centering
  \includegraphics[width=0.8\textwidth]{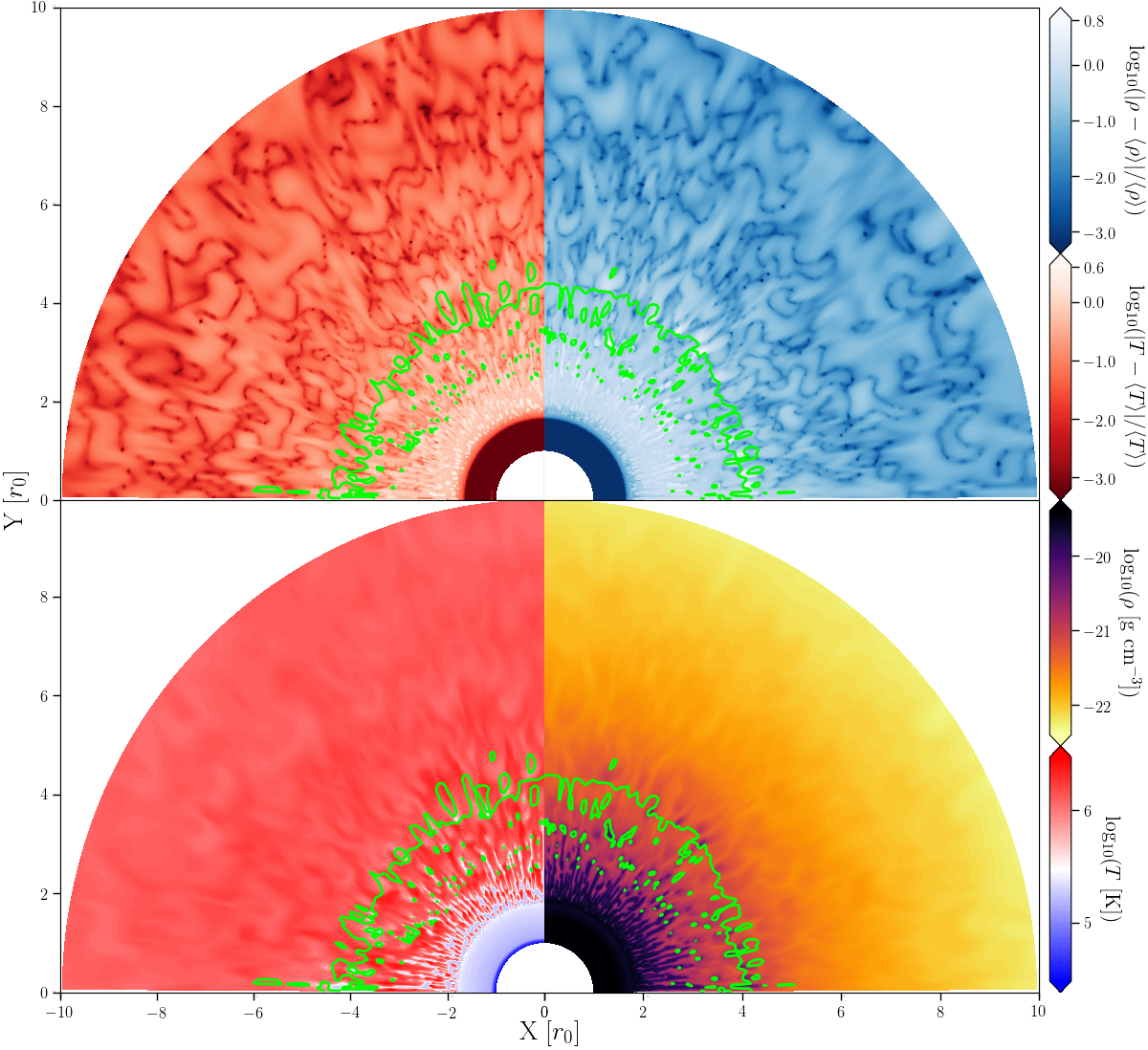}
  \caption{2-D version of model~B. 
  {\it Bottom Panels:} Log-scale maps of $T$ (left) and $\rho$ (right) in cgs units.
  {\it Top Panels:}   The relative difference of $T$ (left)
   and $\rho$ (right) computed using time averaged values. Bright green contours display sonic surfaces.  
   }
  \label{fig:2d-density-contour}
\end{figure*}

Model~D-x1 seems similar to model~A-x1, insomuch as it appears to reach 
a steady state, but model~D-x1 hosts fluctuations at the level of $ \lesssim10\%$.  
This model is also a clear example of a solution that follows 
the S-curve backwards in the ($T$--$\Xi$)-plane
(see the upper right panel of Fig.~\ref{fig:1d-summary-phase}); 
it occupies the same region of the phase diagram as models B and C 
without becoming noticeably unsteady. 
Our analysis of the time scales shows that although $\tau_{\rm TI}$  
is small compared to $\tau_{\rm s}$,
the perturbations that are triggered by
grid scale noise do not stay 
in the TI zone long enough to grow beyond the $10\%$ level in model~D-x1. 
However, model~D type solutions are unstable 
at both higher resolution and to linear 
perturbations inserted by hand, in contrast to model~A type solutions.  
Indeed, we checked that adding random perturbations 
with $10^{-3} \lesssim \delta \rho/\rho_0 \lesssim 10^{-1}$ at $r_0$ 
cause model~D-x1 to behave similarly to model~C-x1.
\par


\subsection{Clumpy winds in 2-D}
We conclude the presentation of our results with one example 
of a 2-D simulation, the counterpart to model~B. 
The bottom panels in Fig.~\ref{fig:2d-density-contour} shows the temperature and density in cgs units, 
whereas the top panels show the relative difference between these quantities and their time-averaged values 
($\langle T \rangle$ and $\langle \rho \rangle$) at a given location for a snapshot near the end of the simulation.
Notice that the flow is very clumpy at $r\simeq 2~r_0$ 
but appears quite smooth at larger radii where the density is overall smaller and clumps have begun to expand.  This is apparent from the density profiles in our 1-D simulations, although not altogether obvious.  The maximum density/temperature contrast is never more than 10. \par

Compared to model~B in 1-D, we found a 
significant scatter of the 2-D wind profiles at a given radius, which reflects the loss of spherical symmetry.  
The amplitude of the scatter (i.e., the variability in the $\theta$ direction) 
is of the same order as the variation in the radial direction.  
This in turn shows that the perturbations 
grow to a similar level in both directions.\par

\section{Discussion} \label{sec:discussion}
In this study, we identified a finite parameter space for which 
a thermally driven wind is clumpy. 
The origin of the clumps in our X-ray irradiated outflows 
is very similar to that of clumpy accretion flows studied 
by \cite{Barai12} and MP13: quite simply, small perturbations 
are allowed to grow due to TI.  However, we found that because 
the flow enters the TI zone near the cold phase, 
TI mainly serves to raise the temperature of perturbations.  
Therefore, the clumpiness is a result of the separation 
of heated layers of gas from the cold, dense layers near the base of 
the flow rather than the formation of dense clumps within 
a more tenuous background plasma.  
Importantly, this formation of heated layers from TI requires 
that a perturbation can stay in the TI zone long enough 
for it to become nonlinear and
that stretching due to radial flow acceleration does not stabilize the growth of such layers.   

These two requirements represent necessary conditions 
for TI to operate in dynamical flows and are due to 
velocity gradients in the wind that are not accounted 
for in the classical theory of TI.  We similarly identified 
another necessary condition that distinguishes dynamical TI 
from local TI, this one arising from pressure gradients: 
$\Xi$ must decrease once gas enters a TI zone (see \S\ref{sec:clumpyruns}).
Because $\Xi$ is the ratio of two pressures, both of which 
are decreasing functions of radius, $\Xi$ could 
be a non-monotonic function of radius. 
Specifically, the models here have $\Xi \propto 1/(r^2 p)$, and even though $p$ decreases with radius, 
this decrease could be slower than $1/r^2$ and $\Xi$ can decrease downstream. 

To build intuition for when to expect a clumpy versus smooth outflow,
we could just consider the geometric effects involved in the problem.
For example, let us assume that the radiation flux scales as
$r^{-q}$, where $q$ is a constant. 
We then find the following trend: the more $q$ is less 
than $2$, the more solutions resemble those from D17, where the wind 
is steady and monotonic  
and $\Xi$ shows no `back tracking'.
For $2.0\lesssim q \lesssim 2.5$, the wind is unsteady and the range of $\Xi$ and $T$ 
is reduced compared to the cases presented here. For $q\gtrsim2.5$, 
the flux drops so fast that the heating is very weak. Consequently,
$\Xi$ and $T$ are never large enough for the solution to even approach 
the TI region and as a result the flow is smooth.

We have explored many other test runs, e.g., with line driving turned on or using a different SED.  
We found that clumpy outflows develop as in the examples shown here but for somewhat different HEP.
We will report on the results from these simulations in a follow-up paper.
\par 
 
\acknowledgments
Support for Program number HST-AR-14579.001-A was provided by NASA through a grant from the Space Telescope Science Institute, which is operated by the Association of Universities for Research in Astronomy, Incorporated, under NASA contract NAS5-26555. This work also was supported by NASA under ATP grant NNX14AK44.\par 


\begin{thebibliography}{}

\bibitem[Balbus(1986)]{Balbus86} Balbus, S.~A.\ 1986, \apjl, 303, L79

\bibitem[Balbus(1995)]{Balbus95} Balbus, S.~A.\ 1995, The Physics of the Interstellar Medium and Intergalactic Medium, 328

\bibitem[Barai et al.(2012)]{Barai12} Barai, P., Proga, D., \& Nagamine, K.\ 2012, \mnras, 424, 728

\bibitem[Crenshaw et al.(1999)]{Cetal99} Crenshaw, D.~M., Kraemer, S.~B., Boggess, A., et al.\ 1999, \apj, 516, 750


\bibitem[Davidson, \& Netzer(1979)]{DN79} Davidson, K., \& Netzer, H.\ 1979, Reviews of Modern Physics, 51, 715

\bibitem[Dyda et al.(2017)]{Detal17} Dyda, S., Dannen, R., Waters, T., \& Proga, D.\ 2017, \mnras, 467, 4161 (D17)

\bibitem[Ebrero et al.(2016)]{Eetal16} Ebrero, J., Kriss, G.~A., Kaastra, J.~S., et al.\ 2016, \aap, 586, A72

\bibitem[Elvis(2017)]{Elvis17} Elvis, M.\ 2017, \apj, 847, 56

\bibitem[Field(1965)]{Field65} Field, G.~B.\ 1965, \apj, 142, 531 

\bibitem[Fu et al.(2017)]{Fetal17} Fu, X.-D., Zhang, S.-N., Sun, W., et al.\ 2017, Research in Astronomy and Astrophysics, 17, 095

\bibitem[Gaspari et al.(2013)]{GRO13} Gaspari, M., Ruszkowski, M., Oh, S.~P., et al.\ 2013, Snowcluster 2013, Physics of Galaxy Clusters, 88

\bibitem[Gabel et al.(2003)]{Getal03} Gabel, J.~R., Crenshaw, D.~M., Kraemer, S.~B., et al.\ 2003, \apj, 583, 178


\bibitem[Higginbottom \& Proga(2015)]{HP15} Higginbottom, N., \& Proga, D.\ 2015, \apj, 807, 107


\bibitem[Kallman \& McCray(1982)]{KMc82} Kallman, T.~R., \& McCray, R.\ 1982, \apjs, 50, 263 

\bibitem[Krolik \& Vrtilek(1984)]{KV84} Krolik, J.~H., \& Vrtilek, J.~M.\ 1984, \apj, 279, 521

\bibitem[Krolik(1999)]{K99} Krolik, J.~H.\ 1999, Active galactic nuclei: from the central black hole to the galactic environment, Julian H.~Krolik.~Princeton, N.~J.~: Princeton University Press, c1999.

\bibitem[Longinotti et al.(2013)]{Letal13} Longinotti, A.~L., Krongold, Y., Kriss, G.~A., et al.\ 2013, \apj, 766, 104

\bibitem[Luketic et al.(2010)]{Letal10} Luketic, S., Proga, D., Kallman, T.~R., et al.\ 2010, \apj, 719, 515

\bibitem[McCourt et al.(2012)]{MSQ12} McCourt, M., Sharma, P., Quataert, E., et al.\ 2012, \mnras, 419, 3319

\bibitem[Mehdipour et al.(2015)]{Metal15} Mehdipour, M., Kaastra, J.~S., Kriss, G.~A., et al.\ 2015, \aap, 575, A22

\bibitem[Mehdipour et al.(2017)]{Metal17} Mehdipour, M., Kaastra, J.~S., Kriss, G.~A., et al.\ 2017, \aap, 607, A28

\bibitem[Mo{\'s}cibrodzka \& Proga(2013)]{MP13} Mo{\'s}cibrodzka, M., \& Proga, D.\ 2013, \apj, 767, 156 (MP13)

\bibitem[Nayakshin(2014)]{Nayakshin14} Nayakshin, S.\ 2014, \mnras, 437, 2404

\bibitem[Proga \& Kallman(2002)]{PK02} Proga, D., \& Kallman, T.~R.\ 2002, \apj, 565, 455

\bibitem[Sharma et al.(2012)]{SMQ12} Sharma, P., McCourt, M., Quataert, E., et al.\ 2012, \mnras, 420, 3174

\bibitem[Shields \& Hamann(1997)]{SH97} Shields, J.~C., \& Hamann, F.\ 1997, \apj, 481, 752

\bibitem[Shlosman et al.(1985)]{SVS85} Shlosman, I., Vitello, P.~A., \& Shaviv, G.\ 1985, \apj, 294, 96

\bibitem[Stone et al.(in prep)]{Stone20} Stone, J. M., Tomida, K., White, C. J., \& Felker, K. G., in preparation, 2020

\bibitem[Takeuchi et al.(2013)]{TOM13} Takeuchi, S., Ohsuga, K., \& Mineshige, S.\ 2013, \pasj, 65, 88

\bibitem[Waters \& Proga(2018)]{WP18} Waters, T., \& Proga, D.\ 2018, \mnras, 481, 2628

\bibitem[Waters \& Proga(2019)]{WP19} Waters, T., \& Proga, D.\ 2019, \apj, 875, 158

\bibitem[Woods et al.(1996)]{Wetal96} Woods, D.~T., Klein, R.~I., Castor, J.~I., et al.\ 1996, \apj, 461, 767

\end{thebibliography}
\end{document}